\documentclass[pra,preprint,aps]{revtex4}
\usepackage{graphics}

\begin{document}

\title{OBLIQUE SURFACE WAVES ON A PAIR OF PLANAR PERIODIC SLOTTED WAVEGUIDES.}

\author{C. Tannous}
\email{tannous@univ-brest.fr}
\altaffiliation{Present address: Laboratoire de Magnétisme de Bretagne, UPRES A CNRS 6135,
 Université de Bretagne Occidentale, BP: 809 Brest CEDEX, 29285 FRANCE}
\affiliation{TRLabs, Suite 108, 15 Innovation Boulevard Saskatoon SK, S7N 2X8, Canada}
\author{R. Lahlou and M. Amram}
\affiliation{Département de Génie physique, Ecole Polytechnique de Montréal 
C.P. 6079, Succursale A, Montréal, PQ, H3C 3A7, Canada}
\date{March 16, 2001}

\begin{abstract}
The dispersion relation and mode amplitudes of oblique surface
waves propagating on an acoustic double comb filter are obtained
with a method based on the calculus of residues. We obtain a
better agreement (below 480 Hz) between theoretical predictions
and measurements reported previously when the filter was being
supposed to be made of a single comb structure.


\end{abstract}

\maketitle

\section{Introduction}

The behavior of a slow wave filter made of a pair of planar
periodic waveguides subjected to low frequency acoustic waves
incident upon the aperture separating the waveguides has been
investigated theoretically and experimentally for its potential
use in acoustic filtering devices \cite{lahlou}. Each waveguide has a comb
structure consisting of a periodic array of blades perpendicular
to a base plane (Figure 1).\\

Using a mathematical model borrowed from the study of electrical
filters, a filter having the same geometric structure of a single
comb waveguide has been analyzed previously [1]. The dispersion
relation, amplitude and phase as functions of frequency and wave
number were derived and compared to experiment. In this work, we
extend our previous theoretical results and consider the actual
nature of the filter consisting of the two waveguides facing each
other. We derive the dispersion relation and
reflection (transmission) coefficients of surface waves propagating
along any oblique wave number in the plane parallel to the comb
structure base planes. \\

Our calculations are based on a weak-coupling approximation and
in the limit of large distance separating the two structures. 
This means the separation is much
larger than the inter-blade distance. The blades are supposed to
have a vanishingly small thickness and we neglect possible
reflections from the planar base affecting the propagating modes,
by direct analogy with the electromagnetic case [2]. This is
equivalent to assuming a slot depth large with respect to the
inverse lowest attenuation of the structure [2].\\

Our work is organized as follows: In section II, we discuss the geometry,
propagating modes dispersion and amplitude relation for the
surface waves. Section III covers the comparison with the
experimental results and the conclusion is in Section IV.

\section{DISPERSION RELATION, MODES AND AMPLITUDES}

Periodic arrays of slotted waveguides stacked to form a
rectangular [3] or prismatic [4] structure are good candidates
for reducing environmental noise (0.1 to 2 kHz). Their
properties have been analyzed theoretically and experimentally
[1, 3, 4] such as their reflection scattering of sound waves
harmful to the general population living near highways or other
sources of damaging sources of low frequency noise. It is
important to understand how these structures absorb, reflect,
transmit or phase delay the incoming sound waves reaching them
with arbitrary time dependent angles. For the rectangular
structure, we have already undertaken such study from the
experimental point of view as well as from the theoretical one.
In this work, we set out to investigate a new type of structure
introduced in detail in Ref. 1 theoretically and experimentally
(Fig. 1).\\

 We have studied the dispersion relation of acoustic
waves impinging on the structure at an arbitrary fixed angle in
the base plane, and measured the sound reflection and
transmission with respect to the incident angle. Our prior
theoretical investigation took account of a single comb structure
only. Here we extend it and deal with a symmetrical weakly-
coupled double comb structure [5] in the limit $\frac{b}{d} \gg 1$ where b
is half the distance between the tip of the blades belonging to
each of the waveguides and d is the inter-blade distance in any
waveguide (Fig. 1).

\begin{figure}[htbp]
\begin{center}
\scalebox{0.5}{\includegraphics*{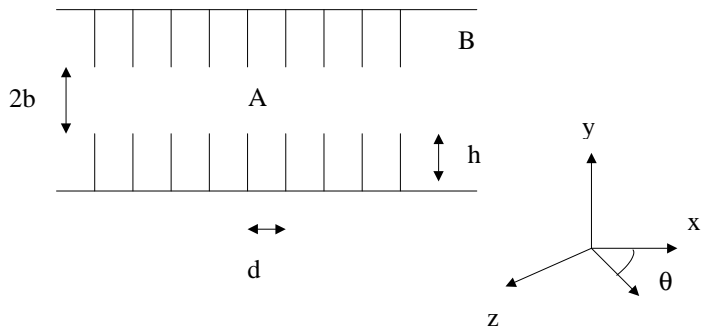}}
\caption{Geometry of the double comb structure waveguide.}
\end{center}
\label{fig1}
\end{figure}

Following our notation [1], we write for the acoustic fields in
region A (free space) keeping the symmetric modes only:

\begin{equation}
\Phi_{A}(x,y,z)= \sum_{n=-\infty}^{\infty}A_ne^{-j\beta_{n}x-j\tau z} \cosh(\alpha_{n}y)
\end{equation}

where $\beta_{n}$ and $\tau$ are the propagation constants along x and z 
and $\alpha_{n}$ is the attenuation constant along y. 
The propagation constant $\beta_{0}$
defining the fundamental mode is determined from the propagation
geometry (Fig. 1 of [6]). It is equal to $\frac{\tau}{\tan(\theta)}$ 
where $\theta$ is the
angle, the surface wave vector makes with the x-axis
[Fig. 1]. The surface wave has a smaller velocity than in true
free space by the ratio $\sqrt{\beta^{2} + {\tau^2}}/k$. 
In region B, the acoustic field in the n-th slot defined by the 
inequalities: $\nu d-\frac{d}{2} \le x \le \nu d + \frac{d}{2}$ is given by:

\begin{equation}
\Phi_{B}^{\nu}(x,y,z)= \sum_{n=-\infty}^{\infty}B_m^{\nu}e^{-j\tau z} \cos(\frac{m\pi x_{\nu}}{d})\cos[\gamma_{m}(y+b+h)]
\end{equation}

The coefficients $B_m^{\nu}$ are determined with the help of Floquet's
[7] theorem $B_m^{\nu}= B_{m} e^{-j\nu \beta_{0} d}$ and the 
abscissae $x_{\nu}$ are equal to x-($\nu$-1/2)d.
In order to find the dispersion equation of the surface waves and
the coefficients $A_{m}, B_{m}$, we will proceed as we did in our
previous work following the approach pioneered by Whitehead [7]
and Hurd [2]. It consists of writing the equations of continuity
for the fields $\Phi_{A} \mbox{and} \hspace{1mm} \Phi_{B}$ and their derivatives along the vertical
y axis on the boundaries $y = \pm b$. These equations are considered
as originating from Cauchy's theorem of residues for a
meromorphic function $f(w)$ taken along some contour and the
contribution of each pole is identified with the contribution of
some corresponding mode. The contour and $f(w)$ should be such that
the presumed theorem of residues is satisfied. Moreover, the
asymptotic behavior of $f(w)$ is tailored by the underlying
physical problem and is basically dictated by the scattering of
the waves by the edges of the blades [7].
We obtain the following meromorphic function $f(w)$ of the complex
variable $w$:

\begin{equation}
f(w)=- \frac{d B_{0} \gamma_{0} e^{-j \gamma_{0}h} }{[e^{-j \beta_{0}d}-1]}
(\frac{j \gamma_{0} - \alpha_{0}}{w - \alpha_{0}} )
\frac{\prod_{1}(w)}{\prod_{1}(j \gamma_{0})}
\frac{\prod_{2}(j \gamma_{0})}{\prod_{2}(w)}
\exp[(j \gamma_{0} -w)\frac{d \hspace{1mm} \ln(2)}{\pi}]
\end{equation}

where $\prod_{1}(w) \hspace{0.5mm} \mbox{and} \prod_{2}(w) $ are the following infinite products:

\begin{equation}
\mbox{$\prod_{1}(w)$} = \prod_{p=1}^{\infty}(w-j \gamma_{p})(-\frac{d}{p \pi})e^{\frac{dw}{p\pi}}
\end{equation}

and:

\begin{equation}
\mbox{$\prod_{2}(w)$} = \prod_{p=1}^{\infty}(w- \alpha_{p})(w+ \alpha_{-p})({\frac{d}{2 p \pi}})^{2} e^{\frac{dw}{p\pi}}
\end{equation}

The propagation constants $\gamma_{m}$ along y, are given by:

\begin{equation}
\gamma_{m}^{2}=k^{2}-\tau^{2}-({\frac{m\pi}{d}})^{2} \hspace{3cm} \mbox{ with $m$=0,1...}
\end{equation}

In order to derive the dispersion relation, we form the ratio:

\begin{equation}
\frac{f(-j\gamma_{0})}{f( j \gamma_{0})}=-e^{2 j \gamma_{0} h }
\end{equation}

Taking the logarithm and using trigonometric identities [Ref. 2], we obtain:

\begin{eqnarray}
&& \gamma_{0} h-\gamma_{0} \frac{d \hspace{1mm} \ln(2)}{\pi} = 
\frac{\pi}{2} -\sin^{-1}(\frac{\gamma_{0}}{\beta_{0}})
+ \sum_{p=1}^{\infty}[\tan^{-1}(\frac{\gamma_{0}}{\alpha_{-p}})
+ \frac{d \hspace{1mm} \gamma_{0} }{2\pi p}]
+ \sum_{p=1}^{\infty}[\tan^{-1}(\frac{\gamma_{0}}{|\gamma_{p}|})
- \frac{d \hspace{1mm} \gamma_{0} }{p \pi}]  \nonumber \\
&& \hspace{5cm} - \sum_{p=1}^{\infty}[\sin^{-1}(\frac{\gamma_{0}}{\beta_{p}})
- \frac{d \hspace{1mm} \gamma_{0} }{2\pi p}]
\end{eqnarray}

This equation is the same as that obtained by Hougardy and Hansen
\cite {hougardy} who treated a single comb structure from the electromagnetic
point of view. Here, we are dealing with the weak coupling
symmetric case limit and with the additional simplifying
assumptions: $\frac{b}{d} \gg 1, \alpha_{0}b \gg 1 \hspace{1mm}
 \mbox{and} \hspace{1mm} \alpha_{-p} \sim \beta_{-p}$. We find that the
dispersion relation is essentially the same as in the case of a
single comb structure. The double comb structure simply behaves
as a single one from the dispersion relation point of view. This
justifies our assumptions in Ref. 1 where we found very good
agreement between theory and experiment up to frequencies on the
order of 400 Hz. Nevertheless this is not true for
reflection (transmission) coefficients of the single/double comb
structures as discussed below.\\

In order to calculate the mode amplitudes and obtain from them
the reflection (transmission) coefficients of the structure, we use
the residue of $f(w)$ at $w=\alpha_{n}$:

\begin{equation}
Res{[f(w)]}_{w=\alpha_{n}}=A_{n}\beta_{n}e^{j \beta_{n}d/2}\cosh(\alpha_{n}b)
\end{equation}

to obtain (n=0, 1, 2...):

\begin{equation}
\frac{|A_{n}|}{|B_{0}|}=\frac{d \gamma_{0} e^{\alpha_{n}b} }{16 \pi \cosh(\alpha_{n}b)}
\frac{|\alpha_{n}+\alpha_{0}|}{|\alpha_{n}\beta_{n}|}
\frac{|\alpha_{n}+\alpha_{1}| \hspace{1mm}|\alpha_{n}-\alpha_{-1}|}{| \alpha_{n}+j\gamma_{1}|}
\frac{ \Gamma[2+\frac{d \alpha_{n}}{\pi}] \hspace{1mm} \exp(-\frac{\alpha_{n}d \hspace{1mm}\ln(2)}{\pi})}{\Gamma[2+\frac{d}{2 \pi}(\alpha_{n}+\beta_{0})]
 \hspace{1mm} \Gamma[2+\frac{d}{2 \pi}(\alpha_{n}-\beta_{0})]}
\end{equation}

where $\Gamma$ stands for the Euler Gamma function. For negative values
of n, it suffices to change $\alpha_{n}$ into $-\alpha_{n}$ in the above expression.
Let us note that when the separation 2b between the two parts of
the structure, becomes very large we recover exactly the
expression found by Hougardy and Hansen \cite {hougardy} corresponding to
a single comb structure.\\

In order to calculate the $B_{n}$ coefficients, we use:

\begin{equation}
f(-j \gamma_{n})=\frac{d}{2} \frac{B_{n} \gamma_{n} e^{j \gamma_{n} h}}
{[{(-)}^{n} \exp(-j\beta_{0} d) - 1]}
\end{equation}

and the definition (3) of $f(w)$ to obtain:

\begin{equation}
\frac{ |B_{n}| }{ |B_{0}| } = \frac{ 2 \gamma_{0} \epsilon }{| \gamma_{n}e^{ j \gamma_{n} h }| }
\frac{|j \gamma_{0} - \alpha_{0}| }{|j \gamma_{n} + \alpha_{0}|}
\frac{ |\mbox{ $\prod_{1}$}(-j\gamma_{n})|}{| \mbox{ $\prod_{2}$ }(-j \gamma_{n})|}
\frac{ \sin( \frac{\beta_{0}d}{2}  ) }{ ( \frac{\beta_{0}d}{2} )  } 
\exp( \frac{ j \gamma_{n} d \hspace{1mm} \ln(2)  } {\pi}  )
\end{equation}

where $\epsilon=1$ for n even, and $\epsilon=\frac{1}{|\tan(\frac{\beta_{0}d)}{2})|}$
 for n odd.\\

Let us note that the $B_{n}$ coefficients are the same as those
obtained by Hougardy and Hansen \cite {hougardy} reflecting the fact, the weak-
coupling approximation affects in a different way the $A_{n}$ and the
$B_{n}$ coefficients. This has important implications on our
measurements of the amplitude profile.

\section{COMPARISON WITH EXPERIMENT}

In our previous work, we derived the dispersion relation,
transmission and reflection coefficients and found excellent
agreement between the single comb structure theory and experiment
up to 400 Hz \cite {lahlou}. This work shows that a weak coupling between
two comb structures does not affect the surface wave dispersion
relations and the $B_{n}$ amplitude coefficients but it does affect
the $A_{n}$ amplitude coefficients.\\

We are going to evaluate how our theory modifies the amplitude
ratio $ \frac{|A_{0}|}{|B_{0}|}$ associated with the fundamental
 mode (n=0) in relation (10) compared with same given in Lahlou et al. \cite {lahlou}.
 The double comb over single comb structure ratio of the two
 expressions is given by:

\begin{equation}
F(\theta) = \frac{e^{\alpha_{0}b}}{2 \cosh(\alpha_{0}b)}
\end{equation}

For a given frequency and a given incident angle $\theta$ we solve the
dispersion relation given by equation (8), obtain the propagation
factor $\alpha_{0}$ and use it in (13). The corrections F($\theta$) in 
dB are plotted versus $\theta$ in the interval [1, 80] degrees
for various frequencies [400-600 Hz]
in Fig.2. 

\begin{figure}[htbp]
\begin{center}
\scalebox{0.5}{\includegraphics*{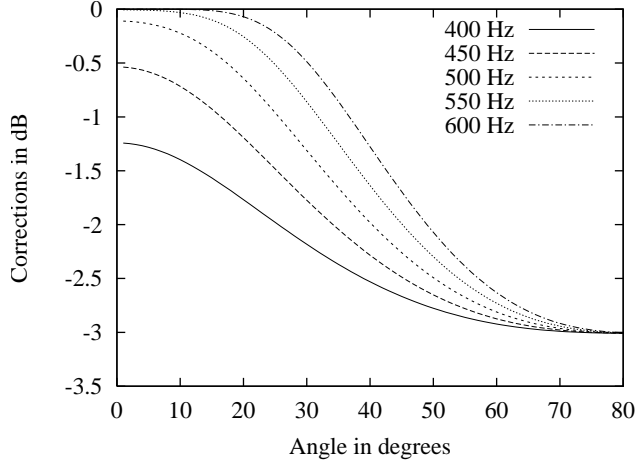}}
\caption{Corrections $F(\theta)$ to the fundamental mode amplitude
ratio $\frac{|A_{0}|}{|B_{0}|}$ with the following values (from the experimental
setup) b=0.0125 m, d=0.05 m, h=0.112 m. The corrections calculated
from $10 \hspace{1mm} log_{10}( \frac{F(\theta=0)}{F(\theta)} )$ are evaluated as a function of the incident angle $\theta$ at a fixed frequency varying from 400 to 600 Hz by steps of 50 Hz.}
\end{center}
\label{fig2}
\end{figure}

The correction comprised between 0 and -3 dB is small
for higher frequencies and small incident angles. It decreases
rapidly for angles larger than 10 to 20 degrees and by a larger amount
for higher frequencies. A comparison to the experimental data
reveals that the correction is pronounced mostly at higher
frequency (336 Hz) and for the largest angle of incidence (47 degrees).
For the highest experimental frequencies (480 and 496 Hz), the
correction introduces more disagreement between the experimental
and theoretical single comb structure theory. This behavior may
be explained by the fact that there are several sources of errors
associated with the measurements at these higher frequencies.

\section{CONCLUSION}

We have developed a weak coupling theory based on the calculus of
residues in order to model the oblique propagation of acoustic
waves propagating through a slow wave filter made of a pair of
comb structured waveguides separated by a distance that is large
with respect to the inter-blade distance. The correction arising
from the symmetrical coupling between the two waveguides has been
evaluated and shown to improve slightly the agreement between the
theoretical and the experimental values of Lahlou et al. 
\cite {lahlou} being at the most 3 dB for the largest frequency
 and angle evaluated. Those results
show that the approximation taken in our previous investigation
is quite acceptable and that the new theory does not bring
substantial additional accuracy to our previous single comb
structure model. Our studies of the strong coupling case ($b < d$)
being mathematically much more complicated, and intended for
improving the agreement between the theoretical results and the
experimental ones at the higher frequencies are in progress and
will be reported in the near future.

\end{document}